\documentclass[aps,amsmath,amssymb,showpacs,twocolumn,floats,epsf]{revtex4}
\usepackage{graphicx,morefloats,slashed}
\usepackage{subfigure}
\usepackage{bm}
\usepackage{color}

\DeclareMathAlphabet{\mathpzc}{OT1}{pzc}{m}{it} 

\begin{document}
\title{High-Chern-number bands and tunable Dirac cones in $\beta$-graphyne }
\author{Guido van Miert}
\author{Cristiane Morais Smith}
\author{Vladimir  Juri\v ci\'c}
\affiliation{Institute for Theoretical Physics, Centre for Extreme Matter and Emergent Phenomena, Utrecht University, Leuvenlaan 4, 3584 CE Utrecht, The Netherlands}

\begin{abstract}
{Graphynes represent an emerging family of carbon allotropes that recently attracted much interest due to the tunability of the Dirac cones in the band structure. Here, we show that the spin-orbit couplings in $\beta$-graphyne could produce various effects related to the topological properties of its electronic bands. Intrinsic spin-orbit coupling yields high- and tunable Chern-number bands, which may host both topological and Chern insulators, in the presence and absence of time-reversal symmetry, respectively. Furthermore, Rashba spin-orbit coupling can be used to control the position and the number of Dirac cones in the Brillouin zone. These findings suggest that spin-orbit-related physics in $\beta$-graphyne is very rich, and, in particular, that this system could provide a platform for the realization of a two-dimensional material with tunable topological properties.
}
\end{abstract}


\maketitle

{\it Introduction.} Since its synthesis in 2004, graphene has attracted enormous attention due to its unusual properties, and has provided a new paradigm for Dirac materials in condensed-matter physics \cite{RMP-CastroNeto2009}. Unconventional electronic properties in this material arise as a consequence of the pseudo-relativistic nature of its low-energy quasiparticles.
Furthermore, graphene represents a platform for the first proposed time-reversal invariant topological insulator \cite{Kane-Mele-PRL2005}, which, however, has not been realized experimentally due to a weak spin-orbit coupling (SOC). Since then, achieving strong SOC in Dirac materials has been one of the goals that motivated the search for alternatives to graphene, and promising candidates have been recently proposed, including self-assembled honeycomb arrays of heavy atom semiconducting nanocrystals, such as Pb and Se \cite{MoraisSmith-PRX2014}, patterned quantum dots \cite{Pellegrini}, as well as  molecular graphene \cite{manoharan}.
An important new emerging class of two-dimensional (2D) carbon allotropes consists of graphynes, among which the most studied members include $\alpha-$, $\beta-$, and $\gamma-$graphynes  \cite{graphyne-review2014}.
They have not been synthesized yet, in contrast to the structurally similar graphdiyne, which has been recently fabricated \cite{Li2010}, but first steps towards that goal have been achieved \cite{Diedrich2010,Diedrich2012}.

\begin{figure*}[t]
\includegraphics[scale=0.6]{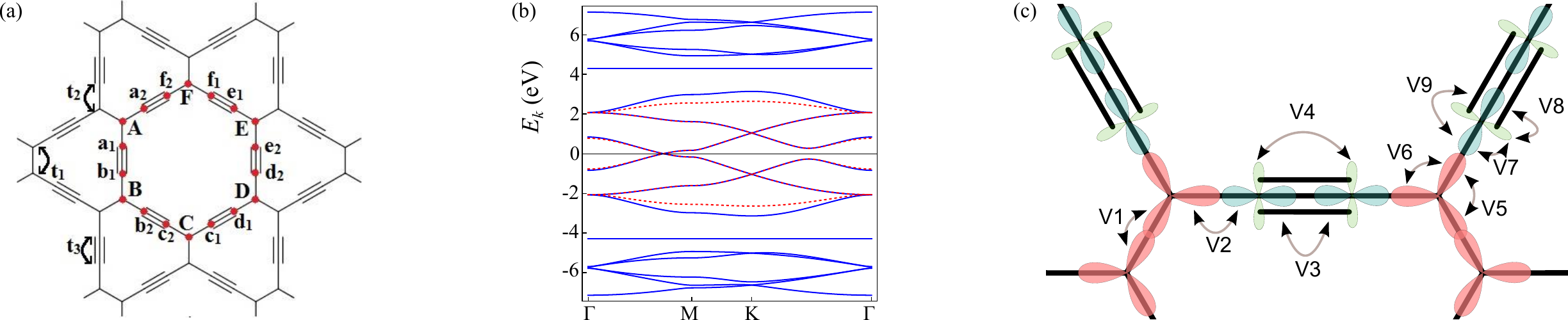}
\caption{(Color online). (a) Lattice structure of $\beta$-graphyne together with NN hopping parameters for $p_z$ orbitals $t_i$ ($i=1,2,3$); (b) Band structures along high-symmetry lines obtained from the full 18-orbital $p_z$-TB model and from the effective six-orbital $p_z$-TB model, shown in blue (solid) and red (dashed) lines, respectively, with Fermi level at zero energy; (c) Hopping parameters ($V_j$, $j=1,..,9$) used in the TB model describing the $\sigma$-orbitals with the red, blue and green colors labeling $sp^2$, $sp$, and $p$ orbitals, respectively.\label{fig:lattices} }
\end{figure*}

Graphynes are allotropes of carbon obtained by inserting a triple bond ($-$C $\equiv$ C$-$) into the graphene structure, see Fig.~\ref{fig:lattices}(a). These 2D allotropes of carbon  were proposed in 1987 \cite{baughman1987}, and their structural, electronic, and mechanical properties have been rather extensively studied  \cite{narita1998,galvao2003,tahara2007,kang2011,yue2012,cranford2011,Kim-Choi2012}. They have recently attracted much interest, especially because of the existence of tunable Dirac cones in their band structure, and in that sense could be even more promising for applications than graphene \cite{malko2012,Huang2013}.
  Features of the Dirac cones may be controlled by chemical reactions \cite{JJZheng2013} and adatoms  \cite{ zhengJChemPhys2013, HeJPhysChemC2012,GarrityPRL2013}. Adatoms of heavy elements, such as Indium and Thallium, could also be used to tune SOC in graphynes, as suggested in Ref.\ \cite{franz-prx} for graphene,  which is particularly important in light of inducing and manipulating topological  properties,  critically dependent on its strength.

Motivated by these developments, in this Rapid Communication we study the effects of the SOCs  in $\beta$-graphyne [(Fig.~\ref{fig:lattices}(a)], and show that this material could be advantageous with respect to graphene not only concerning the tunability of the Dirac cones \cite{malko2012}, but also regarding topological properties of its band structure. To do so, we derive  an effective tight-binding (TB) theory that includes both Rashba and intrinsic SOCs. So far, their effects have been only sparsely studied using {\it ab initio} methods  \cite{MZhaoSciRep2013}, but clearly an effective description of the SOCs in graphynes is highly desirable.
Before elaborating on the specific results, we first observe that due to the lattice structure of  $\beta-$graphyne, a minimal effective TB model contains six low-energy $p_z$ orbitals, one on each of the six sites on the vertices within the unit cell. Second, for either of the two types of SOC there are, in fact, two parameters, which we call {\it internal} and {\it external}, describing the corresponding intra- and inter-unit-cell hoppings. In graphene, these two parameters are equal, since intra- and inter-unit-cell bonds therein are equivalent. This fact in conjunction with the larger unit cell (18 atoms in $\beta$-graphyne versus two in graphene) already hints that the SOC-related physics in $\beta$-graphyne could be richer than in graphene.

\begin{figure*}[t]
\centering
\includegraphics[scale=0.55]{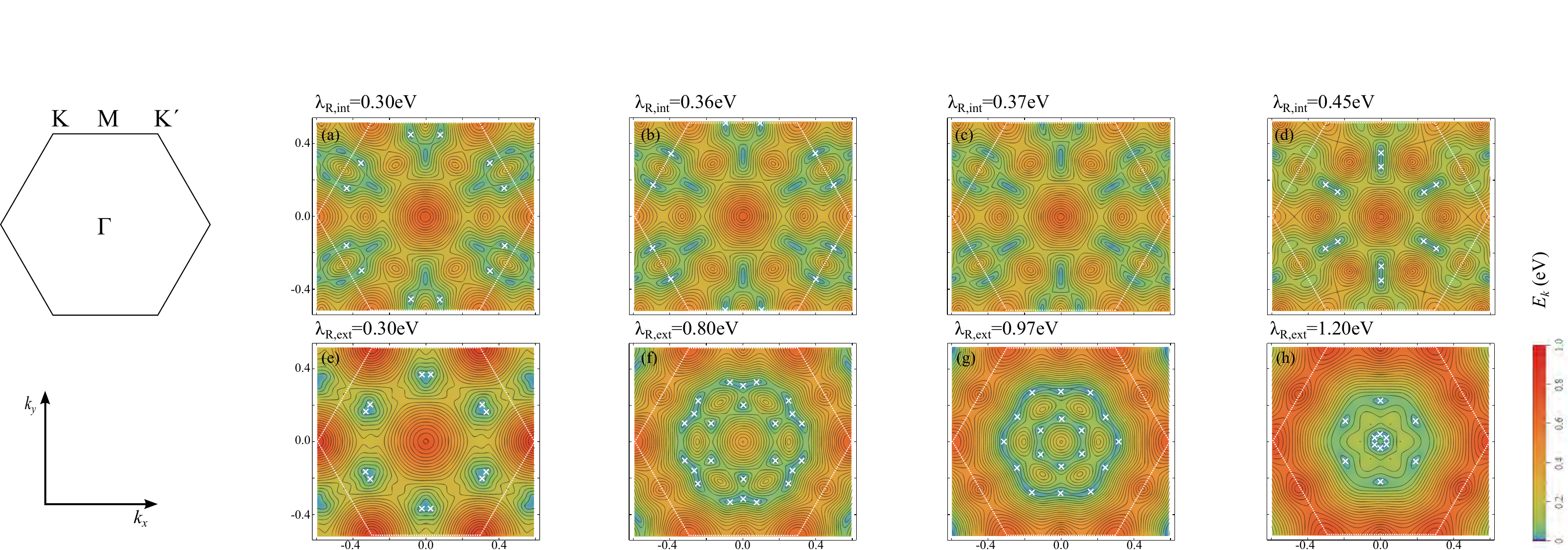}
\caption{(Color online). Band structure in the presence of the Rashba SOC for $t_{\rm ext}/t_{\rm int}=-1.18$.  The panels in the upper (lower) row display the Lifshitz transitions driven by the internal (external) Rashba SOC. The BZ is represented by a white line, the Dirac cones are labeled by white crosses in the plots. The high-symmetry points in the BZ are shown in the upper left panel and momentum is in units with the lattice constant set to one.\label{fig:merging} }
\end{figure*}

\begin{figure*}[t]
\centering
\includegraphics[scale=0.75]{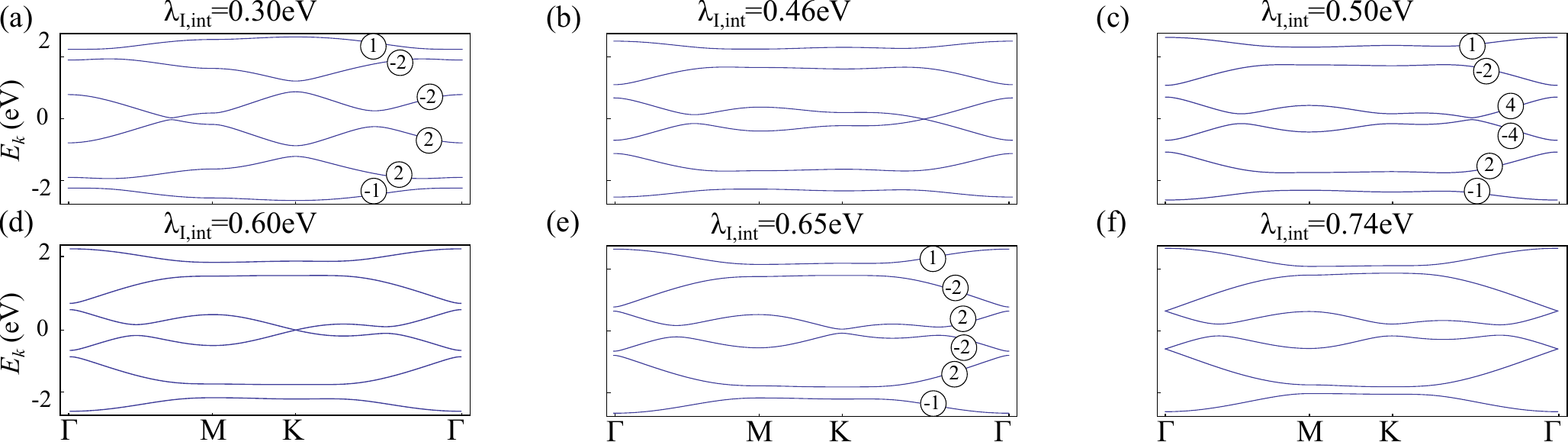}
\caption{(Color online). The six panels display the band structure along high-symmetry lines for $\beta$-graphyne with the internal intrinsic SOC. The numbers in the circles denote the Chern numbers for the respective bands with spin up.\label{fig:chern}}
\end{figure*}

Indeed, this is the case. Without SOCs, there are six  {\it inequivalent}  Dirac cones at the Fermi level along the high-symmetry $\Gamma-M$ line in the Brillouin zone (BZ) [Fig.\ \ref{fig:lattices}(b)].   As the internal Rashba SOC is cranked up, we show that the system undergoes a series of Lifshitz phase transitions in which the Dirac cones split, merge and split again, however, in different directions in the BZ (Fig.\ \ref{fig:merging}).
The effect of the intrinsic SOC, on the other hand, is to produce gapped topologically nontrivial bands with {\it tunable} Chern numbers as high as $C=4$ per spin [Fig.\ \ref{fig:chern}], suggesting that $\beta$-graphyne could provide the ground for realizing both conventional (noninteracting) and fractional topological insulators upon accounting for interactions \cite{FTI-review}. In fact, as we show here, various topologically-insulating  phases may emerge at half-filling both in the presence and in the absence of time-reversal symmetry (TRS), demonstrating the possibility of achieving topological phase transitions in a carbon-based material.

{\it Tight-binding description of the SOC}. Among the three most studied graphynes, $\beta$-graphyne has the most complicated lattice structure. Its unit cell consists of a hexagon, which has one carbon atom located at each vertex, and two carbon atoms connected by an acetylene linkage between each two neighboring vertices, yielding together 18 atoms, see Fig.~\ref{fig:lattices}(a).
For the description of SOC effects, we use a basis that contains two sets of orbitals: $\sigma$ orbitals, consisting of the atomic $s$, $p_x$, and $p_y$ orbitals, as well as $p_z$ orbitals. To obtain an effective Hamiltonian containing only six $p_z$ orbitals at the vertices, we  proceed in two steps: (i) we first integrate out high-energy $\sigma$-orbitals to obtain a Hamiltonian with 18 $p_z$ orbitals at both vertex and edge sites in the unit cell; (ii) in the second step, we eliminate 12 $p_z$ orbitals at the edge sites.

We first analyze the system without SOC, when the $\sigma$ and $p_z$ orbitals are decoupled, and only the latter are relevant for the low-energy description at half-filling.
The corresponding band-structure is displayed in Fig.~\ref{fig:lattices}(b)(blue solid lines). Using step (ii) in the outlined procedure, we obtain an effective Hamiltonian with only six $p_z$ orbitals located at the lattice vertices \cite{zhe}
 \begin{align}
 \label{effb}
&H_{z}^{\rm eff}=t_{\rm int}\sum_{\langle i,j\rangle}[A^\dagger_i\left(B_j+F_j\right)+C^\dagger_i\left(B_j+D_j\right)+E^\dagger_i(D_j\nonumber\\
& +F_j)]
+t_{\rm ext}\sum_{\langle i,j\rangle}[A^\dagger_i D_j+C^\dagger_i F_j+E^\dagger_i B_j]+h.c.
\end{align}
 with $t_{\rm int}=-t_2^2t_3/T$ and $t_{\rm ext}=t_1t_3^2/T$, denoting nearest-neighbor (NN) intra-unit-cell and inter-unit-cell hoppings, respectively, and $T\equiv 2t_2^2+t_3^2$. The hopping parameters for the 18 $p_z$ orbitals are [Fig.\ \ref{fig:lattices}(a)] $t_{1}=-2.00$eV, $t_{2}=-2.70$eV, and $t_{3}=-4.30$eV \cite{zhe}, hence $t_{\rm int}=0.95$eV and $t_{\rm ext}=-1.12$eV.
 It turns out that this Hamiltonian captures the low-energy band-structure at half-filling very well, see Fig.~\ref{fig:lattices}(b) (red dashed lines). Note that the dispersion relation exhibits six Dirac cones at the Fermi level, located on the line $\Gamma-M$. As opposed to graphene and $\alpha-$graphyne, where the cones exhibit a threefold  rotational symmetry, in $\beta-$graphyne the cones are symmetric only under mirror reflection through the normal plane containing the line $\Gamma-M$, and are thus anisotropic \cite{malko2012}.

We now turn to the description of the effects arising from both intrinsic and Rashba SOCs with the Hamiltonian
\begin{equation}\label{SOC-microscopic}
H_{\rm SOC}=H_L+H_E.
\end{equation}
The intrinsic SOC originates from relativistic effects, microscopically described by the Hamiltonian $H_{L}=-f(r){\bm\sigma}\cdot {\bf L}$, where $f$ is a function related to the effective short-ranged nuclear electrostatic potential, ${\bm \sigma}$ is a vector of Pauli matrices, and ${\bf L}$ is the orbital angular momentum. Furthermore, this coupling possesses mirror symmetry through the lattice ($x-y$) plane, represented by the matrix $\sigma_z$ in spin space. This symmetry then allows  for the coupling between $p_{z,\uparrow}$, $p_{x,\downarrow}$, $p_{y,\downarrow}$, and $s_{\downarrow}$ orbitals, odd under this reflection. Analogously, the orbitals $p_{z,\downarrow}$, $p_{x,\uparrow}$, $p_{y,\uparrow}$,  and $s_{\uparrow}$, even under this symmetry, can be coupled.  Moreover, when an external electric field is applied perpendicularly to the plane, the microscopic Hamiltonian includes an extra term
$H_{E}=E z$,
with $E$ the electric field, that couples the $s$ to the $p_z$ orbitals due to the broken mirror symmetry. A linear in $E$ combination of the Hamiltonians $H_L$ and $H_E$ generates the Rashba SOC, whereas the intrinsic SOC is generated exclusively by $H_L$ \cite{supplemental}.  In addition to the $\sigma$ orbitals, in principle, the $d_{xz}$ and $d_{yz}$ orbitals have to be taken into account for SOC \cite{fabian}. However, the $d$ orbitals are expected not to lead to any qualitatively different effects in $\beta-$graphyne, and we do not  consider those hereafter \cite{us-PRB}.

To derive an effective model describing the SOC, we apply the outlined two-step procedure to the Hamiltonian
\begin{equation}\label{fullHam}
H=H^z_0+H^\sigma_0+H^{z,\sigma}_{E}+H^{z,\sigma}_{L}+\left(H^{z,\sigma}_{E}\right)^\dagger+\left(H^{z,\sigma}_{L}\right)^\dagger,
\end{equation}
where $H^{z}_0$ ($H^\sigma_0$) describes hoppings between $p_z$ ($\sigma$) orbitals denoted by $t_i$ ($V_i$)  [see Figs.\ \ref{fig:lattices}(a), \ref{fig:lattices}(c), and Supplemental Material \cite{supplemental}], while
the SOC terms, obtained from  Eq.\ (\ref{SOC-microscopic}), read
\begin{align}\label{SOC-micro}
H^{z,\sigma}_{E}&=\sum_{j,\alpha=1,2}\xi_{sp\alpha}p^\dagger_{z,j\alpha}s_{j\alpha},\nonumber\\
H^{z,\sigma}_{L}&=\sum_{j,\alpha=1,2}\xi_{p\alpha}p^\dagger_{z,j\alpha}\left(-i\sigma_y p_{x,j\alpha}+i\sigma_x p_{y,j\alpha}\right).\nonumber
\end{align}
 Here, $p_{z,i}^\dagger$ creates an electron in a $p_z$ orbital at position $i$, and analogous notation is used for the $p_x$, $p_y$, and $s$ orbitals, while $\alpha=1(2)$ corresponds to the sites at edge (vertex).
 The exact value of the on-site coupling parameters $\xi_{p1}$, $\xi_{p2}$, $\xi_{sp1}$, and $\xi_{sp2}$ may be obtained by fitting the band structure to {\it ab initio} calculations, which, however, have been performed only for $\alpha$-, $\delta-$, and $6,6,12$-, but not for $\beta$-graphyne with intrinsic SOC \cite{JJZheng2013}.  Note that $\xi_{sp1}$ and $\xi_{sp2}$ are both linear in $E$.
 An approximate lower bound for $\xi_{p1}\simeq12.6$meV found in $\alpha$-graphyne may also apply to $\beta$-graphyne since the charge distribution around the acetylene bond  is approximately the same for all graphynes. To obtain an estimate for $\xi_{p2}$, we use that the charge distribution around the vertices in $\beta$-graphyne and graphene are approximately the same, and therefore we expect that their values should be comparable, yielding $\xi_{p2}\simeq2.8$meV \cite{fabian}.  The parameters $\xi_{sp1}$ and $\xi_{sp2}$ should be approximately equal, since they derive from the corresponding matrix elements of the microscopic Hamiltonian $H_E$ for the wavefunctions at the edges and vertices, expected to be comparable, yielding $\xi_{sp1}\approx\xi_{sp2}\simeq10$meV for a  typical $E\simeq0.1$V/nm \cite{fabian}. These estimates are, however, expected to be modified in the presence of heavy adatoms  and strain. The latter should  influence the charge distribution around the edges making it more inhomogeneous, and therefore enhancing the value of the SOC parameters.

  The obtained effective Hamiltonian that contains only six $p_z$ orbitals reads
  \begin{equation}\label{eff-SOC-pz}
  H^{\rm eff}=H_z^{\rm eff}+H_{\rm R}^{\rm eff}+H_{\rm I}^{\rm eff},
  \end{equation}
 with $H_z^{\rm eff}$ given by Eq.\ (\ref{effb}), while the Rashba and the intrinsic SOCs are given, respectively, in terms of NN and next-NN TB Hamiltonians,
 \begin{align}
 \label{intext}
 H_{\rm R}^{\rm eff}&=i \sum_{a,\langle i,j\rangle}\lambda_{R,a}\, p_{z,i}^\dagger \, ({\bm \sigma}\times \hat{{\bf d}}_{ij})\cdot\hat{{\bf z}}\, p_{z,j},\\
\label{intext2}H_{\rm I}^{\rm eff}&=i \sum_{a,\langle\langle i,j\rangle\rangle}\,\lambda_{I,a}\,v_{ij}\,p_{z,i}^\dagger\,\sigma_z\, p_{z,j}.
\end{align}
 Here, the index $a={\rm int}({\rm ext})$ refers to the SOCs effectively described by the intra-(inter-)unit-cell hoppings with the summations also taken correspondingly, $\hat{{\bf d}}_{ij}$ is a unit vector connecting NNs, and $v_{ij}=+(-)$ if the hopping is (anti-)clockwise. The parameters of the six-band effective SOC Hamiltonian (\ref{eff-SOC-pz}), derived from the Hamiltonian (\ref{fullHam}), are given in Table \ref{table1}.

 \begin{table}[t]
\centering
\begin{tabular}{ l || c  }
   & Microscopic parameters \\
  \hline\hline
  $\lambda_{\rm R,int}$ & $-{\tilde T}[2t_3V_3(\sqrt{2}\xi_{p2}\xi_{sp1}+\xi_{p1}\xi_{sp2})+\sqrt{6}t_2V_2\xi_{p1}\xi_{sp1}]/\sqrt{6}$ \\
  \hline
  $\lambda_{\rm R,ext}$ & $2\sqrt{2}t_3^2\xi_{p2}\xi_{sp2}/(3TV_1)$\\
  \hline
  $\lambda_{\rm I,int}$ & $\sqrt{3}V_6t_2^2\xi_{p1}^2/(4TV_2^2)$ \\
  \hline
  $\lambda_{\rm I,ext}$  & $-t_2t_3V_5\xi_{p1}\xi_{p2}/(2TV_1V_2)$\\
  \hline
\end{tabular}
\caption{Effective Rashba and intrinsic SOC parameters for $\beta$-graphyne in Hamiltonians (\ref{intext}) and (\ref{intext2}), respectively, which are
derived using a two-step procedure from the microscopic Hamiltonian containing one $p_z$ and three $\sigma$ orbitals per site of the 18-atom unit cell with the hoppings $t_i$ ($i=1,2,3$) and $V_j$ ($j=1,...,9$) shown in Figs.\ \ref{fig:lattices}(a) and \ref{fig:lattices}(c). Here,  ${\tilde T}\equiv t_2/(TV_2V_3)$, with $T$ defined below Eq.\ (\ref{effb}).}
\label{table1}
\end{table}

{\it Effects of the Rashba SOC: tunable Dirac cones.}
We first neglect the intrinsic SOC and consider the effects of the internal Rashba SOC, $\lambda_{\rm R,int}$, since this term is expected to dominate over the external one in $\beta$-graphyne. This can be understood from the fact that the internal Rashba SOC arises from the hoppings through the acetylene bond. Since these hoppings yield three as many possibilities to flip spin as the ones not involving acetylene bonds, which pertain to the external Rashba SOC, the internal Rashba SOC is thus expected to dominate. Indeed,  estimates for the microscopic SOC parameters, together with expected similar values for the hoppings between $\sigma$ orbitals $V_1\approx V_2\approx V_3(\simeq5-10$eV) \cite{fabian}, yield $\lambda_{\rm R,int}/\lambda_{\rm R,ext}\approx 5.1$. Using the TB Hamiltonian in Eq.\ (\ref{eff-SOC-pz})  for $\lambda_{I,a}=0$, $\lambda_{\rm R,ext}=0$, and small values of the internal Rashba SOC parameters, we find   that each of the six symmetry-related spin-degenerate Dirac cones at the $\Gamma-M$ lines splits into a pair perpendicular to this line  [Fig.\ \ref{fig:merging}(a)]. As this coupling is increased, the pair moves towards the edge of the BZ [Fig.\ \ref{fig:merging}(b)], eventually annihilates at the line connecting the $K$ and $K'$ points with another pair, and a topologically trivial gap then opens up [ Fig.\ \ref{fig:merging}(c)]. Furthermore, at an intermediate value of $\lambda_{\rm R,int}$ we find that a new pair of Dirac cones emerges, but in this case along the $\Gamma-M$ line [Fig.\ \ref{fig:merging}(d)]. It should be stressed that the order of these Lifshitz phase transitions depends on the precise value of the ratio of the hoppings $t_{\rm ext}/t_{\rm int}$. For instance, when its absolute value is slightly larger than $1.18$, the pairs on the $\Gamma-M$ line emerge {\it before} the other pairs annihilate, and, as $\lambda_{\rm R,int}$ is further increased, only the former survive (not displayed). In fact, as shown in Figs.\ \ref{fig:merging}(e)-(h), the external Rashba SOC alone creates precisely the effect previously described, therefore opening up the possibility to control the Lifshitz phase transitions with either of the two types of Rashba SOC allowed here.

{\it Effects of the instrinsic SOC: high-Chern-number bands.} The intrinsic SOC leads to the formation of topologically nontrivial electronic bands in the system. In $\beta$-graphyne internal and external intrinsic SOC turn out to have the opposite sign (Table \ref{table1}), which results in the enhancement of the topological bandgaps as compared to the case when only one of the couplings is present. This can be readily understood from the fact that this sign difference in the six-site effective model arises after integrating out the high-energy orbitals at the edges in the 18-site TB model, which contains both terms with the same sign. Therefore, we will here only consider the effect of the internal intrinsic SOC effectively described by the hopping within the unit cell, $\lambda_{\rm I,int}$ in Hamiltonian (\ref{intext2}). For an infinitesimal value of this coupling, topologically nontrivial bands arise with the corresponding total Chern number at half-filling $C_\uparrow=3$  for spin-up electrons [Fig.\ \ref{fig:chern}(a)], as opposed to graphene, where the intrinsic SOC produces $C_\uparrow=1$.  This value of the total Chern number in $\beta$-graphyne is a result of the following values of Chern numbers  for the six bands $\{-1,2,2,-2,-2,1\}$ \cite{supplemental}. Since the  ${\mathbb Z}_2$ invariant  is given by the parity of the spin Chern number, $\nu=C_{\uparrow}\,\,({\rm mod\,\, 2})$, $\beta$-graphyne is a topologically nontrivial insulator. On the other hand, when TRS is broken, for instance by a magnetic field, the system  turns into a Chern insulator with Chern number $C=3$, implying that the Hall conductivity $\sigma_{\rm Hall}=3e^2/h$, which differs from graphene where under the same circumstances $\sigma_{\rm Hall}=e^2/h$. We notice here in passing that $\beta$-graphyne cannot become topological crystalline insulator, since it possesses time-reversal invariant momenta only at the $\Gamma$ and three symmetry-related $M$ points in the BZ \cite{space-group-class}. Furthermore, at a critical value $\lambda_{\rm I,int}^{\rm crit}=0.46$eV,  the bandgap at the Fermi level closes at the $\Gamma-K$ line [Fig.\ \ref{fig:chern}(b)] and reopens yielding the Chern numbers $\{-1,2,-4,4,-2,1\}$, and $C_\uparrow=-3$ [Fig.\ \ref{fig:chern}(c)]. In addition, for even stronger internal intrinsic SOC, $\lambda_{\rm I,int}=0.$6eV , the bandgap closes at the $K$ and $K'$ points [Fig.\ \ref{fig:chern}(d)], and upon further increase of this coupling the system enters a topologically nontrivial insulating state with Chern numbers $\{-1,2,-2,2,-2,1\}$ [Fig.\ \ref{fig:chern}(e)].  Finally, for $\lambda_{\rm I,int}=0.74$eV, the bandgap closes at the $\Gamma$ point [Fig.\ \ref{fig:chern}(f)], and for even stronger $\lambda_{\rm I,int}$ an insulator with Chern numbers $\{-1,1,-1,1,-1,1\}$ appears.

{\it Discussion and conclusions.} To summarize, we here demonstrated that $\beta$-graphyne exhibits rich behavior due to the conspiracy of its lattice structure with a relatively large unit cell and the effects of the intrinsic and Rashba SOCs. The latter interaction allows for the tuning of the position and the number of  Dirac cones in the band structure therefore opening up the possibility to manipulate the transport properties of the system. Furthermore, we showed that in $\beta$-graphyne the Dirac cones can be located at all high-symmetry points in the BZ and a plethora of Lifshitz phase transitions between semimetallic phases can be implemented.
Finally, recent progress in realizing honeycomb optical lattices exhibiting tunable Dirac cones \cite{nature-esslinger2012}, in conjunction with the possibility to manipulate SOC in these systems \cite{galitski-spielman-nature2013}, make our findings relevant also for ultracold atoms in optical lattices.

 Not only the Rashba SOC induces interesting effects, but the  intrinsic SOC also does so. Indeed, the latter yields topologically nontrivial bands, some of which possess high Chern number. Moreover, when TRS is broken and the SOC is tuned, e.g., by the presence of heavy adatoms such as Bi and Sn, a series of topological phase transitions between different Chern insulators is expected to occur. However, further {\it ab initio} studies are needed to quantitatively  establish the effect of adatoms on the SOC in graphynes.
The topological bandgap closings are  at $\Gamma$, $M$, $K$ as well as at points in between, and this system therefore may interpolate between graphene where topological  phase transitions occur at the $K$ and $K'$ points \cite{Kane-Mele-PRL2005}, and HgTe quantum wells with the topological bandgap closing at the $\Gamma$ point  \cite{BHZ,molenkamp,cristiane-new}.
Furthermore, the interplay of SOC and magnetic field (or any TRS-breaking perturbation) is a rich and fundamentally important problem, as the studies in graphene \cite{wouter2012} and silicene  \cite{nagaosa2014} have shown.  Finally, we hope that our results will boost {\it ab initio} studies of the spin-orbit effects in the graphyne family of carbon allotropes.

We would like to thank Douglas S. Galv\~ ao for fruitful discussions. This work is part of the D-ITP consortium, a program of the Netherlands Organisation for Scientific Research (NWO) that is funded by the Dutch Ministry of Education, Culture and Science (OCW). The authors acknowledge financial support from NWO.

\clearpage
\newpage

\begin{widetext}
\begin{center}

{\bf Supplemental Material for  "High-Chern-number bands and tunable Dirac cones in $\beta$-graphyne"}\\

\vspace{1em}

{Guido van Miert, Cristiane Morais Smith, and Vladimir  Juri\v ci\'c}\\

\vspace{1em}

{\it Institute for Theoretical Physics, Centre for Extreme Matter and Emergent Phenomena, Utrecht University, Leuvenlaan 4, 3584 CE Utrecht, The Netherlands}\\
\end{center}

\begin{center}
{In this supplemental material we provide the details of the derivation of the spin-orbit Hamiltonians within the tight-binding approach. We also explain the calculation of the Chern numbers.}\\
\end{center}

\section{Tight-binding Hamiltonian}
\subsection{Tight-binding  model without SOC}
\begin{figure}[b]
\centering
\includegraphics[width=.85\textwidth]{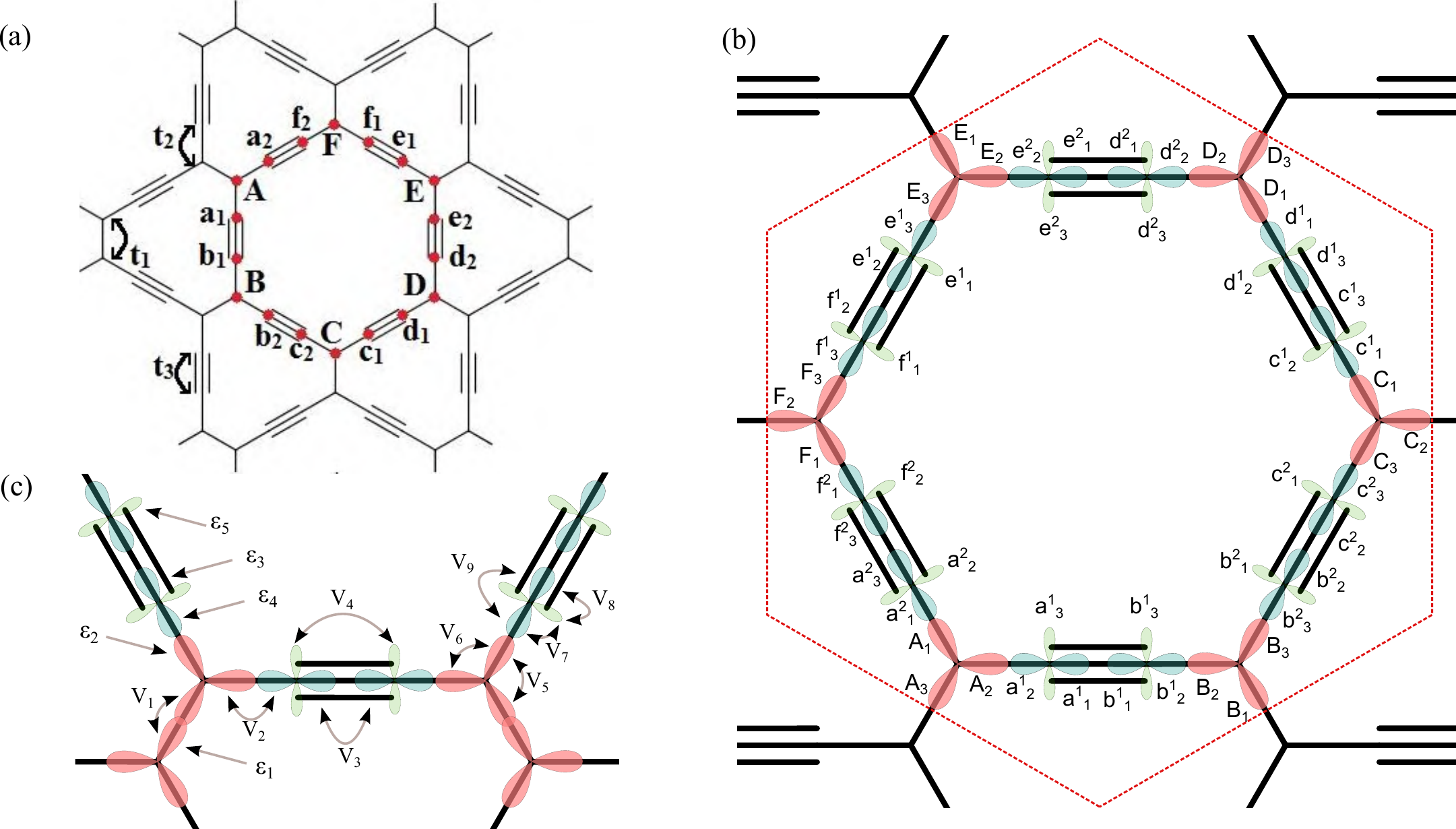}
\caption{(Color online). Lattice structure of $\beta$-graphyne. Orbitals located at the vertices are denoted by  a capital letter, whereas the ones at the  edges are represented by a lower-case letter. (a) The $p_z$ orbitals  are shown, together with the corresponding hopping parameters $t_i$, $i=1,2,3$. The subscript $j=1,2$ further distinguishes these orbitals at the edge sites. (b) Representation of the $\sigma$-orbitals. Here, the subscripts represent the different orbitals, whereas the superscript denotes the lattice position. The $p$ orbitals are oriented in such a way that their positive sides on each lattice site point inwards the  unit cell. (c) Hopping parameters $V_i$, $i=1,\ldots,9$ and on-site energies $\varepsilon_i$, $i=1,\ldots,5$ used in the TB model describing the $\sigma$-orbitals. Note that the $sp^2$, $sp$, and $p$ hybrid orbitals are depicted in red, blue, and green, respectively.}
\label{figa}
\end{figure}
First, we introduce the tight-binding (TB) Hamiltonian in the absence of spin-orbit coupling (SOC).  Since we neglect the spin degree of freedom and the system exhibits reflection symmetry through the $x-y$ plane, it follows that the $p_z$ and $\sigma$-orbitals decouple. As a result, the full Hamiltonian is the sum of two terms,
\begin{align}
H_0&=H^z_0+H^\sigma_0.
\end{align}
The Hamiltonian $H^z_0$ describes the $p_z$ orbitals and reads [see Fig.~\ref{figa}(a)]
\begin{align}
H_0^z&=t_{1}\sum_{\langle i,j\rangle} \left(A^\dagger_i D_j+B^\dagger_i E_j+C^\dagger_i F_j\right)+t_{2}\sum_{\langle i,j\rangle}\left[A^\dagger_i\left(a_{1,j}+a_{2,j}\right)+B^\dagger_i\left(b_{1,j}+b_{2,j}\right)\right.+C^\dagger_i\left(c_{1,j}+c_{2,j}\right)+D^\dagger_i\left(d_{1,j}+d_{2,j}\right)\nonumber\\
&+E^\dagger_i\left(e_{1,j}+e_{2,j}\right)\left.+F^\dagger_i\left(f_{1,j}+f_{2,j}\right)\vphantom{A^\dagger_i}\right]+t_{3}\sum_{\langle i,j\rangle}\left(a^\dagger_{1,i}b_{1,j}+b^\dagger_{2,i}c_{2,j}\right.\left.+c^\dagger_{1,i}d_{1,j}+d^\dagger_{2,i}e_{2,j}+e^\dagger_{1,i}f_{1,j}+f^\dagger_{2,i}a_{2,j}\right)+h.c.\nonumber
\end{align}
This Hamiltonian is relatively simple, since it involves only three hopping parameters $t_i$ [Fig.\ \ref{figa}(a)]. On the other hand, the Hamiltonian describing the $\sigma$ orbitals is more complicated. First of all, we describe the $\sigma$-orbitals in terms of the $sp$, $sp^2$, and $p$ hybrid orbitals, the labeling of which is depicted in Fig.~\ref{figa}(b). Furthermore, we only take into account the onsite energies and hoppings, in addition to the dominant nearest-neighbor (NN) hoppings forming the bonds, see Fig.~\ref{figa}(c). Therefore, we may write
\begin{align}
H^\sigma_0&=H_{\rm \sigma, NN}+H_{\rm \sigma, onsite}.
\end{align}

The Hamiltonian describing the dominant NN hoppings reads
\begin{align}\label{eqn2}
H_{\rm \sigma, NN}&=V_1\sum_{\langle i,j\rangle}\left(A_{3,i}^\dagger D_{3,j}+B_{1,i}^\dagger E_{1,j}+C_{2,i}^\dagger F_{2,j}\right)+V_2\sum_{\langle i,j\rangle}\left( A_{1,i}^\dagger a_{1,j}^2+A_{2,i}^\dagger a^1_{2,j}+B_{2,i}^\dagger b_{2,j}^1+B_{3,i}^\dagger b^2_{3,j} +C_{1,i}^\dagger c^1_{1,j}\right.\nonumber\\
&\left.+C_{3,i}^\dagger c_{3,j}^2+D_{1,i}^\dagger d^1_{1,j}+D_{2,i}^\dagger d_{2,j}^2+E_{2,i}^\dagger e^2_{2,j}+E_{3,i}^\dagger e_{3,j}^1+F_{1,i}^\dagger f_{1,j}^2+F_{3,i}^\dagger f_{3,j}^1\right)+V_3\sum_{\langle i,j\rangle} \left[(a^1_{1,i})^\dagger b^1_{1,j}+(a^2_{3,i})^\dagger f^2_{3,j}\right.\nonumber\\
&\left.+(c^1_{3,i})^\dagger d^1_{3,j}+(c^2_{2,i})^\dagger b^2_{2,j}+(e^1_{2,i})^\dagger f^1_{2,j}+(e^2_{1,i})^\dagger d^2_{1,j}\right]+V_4\sum_{\langle i,j\rangle} \left[(a^1_{3,i})^\dagger b^1_{3,j}+(a^2_{2,i})^\dagger f^2_{2,j}+(c^1_{2,i})^\dagger d^1_{2,j}\right.\nonumber\\
&\left.+(c^2_{1,i})^\dagger b^2_{1,j}+(e^1_{1,i})^\dagger f^1_{1,j}+(e^2_{3,i})^\dagger d^2_{3,j}\right]+h.c.
\end{align}
Similarly, we can write the Hamiltonian describing the on-site energies and hoppings
\begin{align}\label{eqn3}
H_{\rm \sigma, onsite}&=\frac{\varepsilon_1}{2}\sum_{i}\left( A_{3,i}^\dagger A_{3,i}+B_{1,i}^\dagger B_{1,i}+C_{2,i}^\dagger C_{2,i}+D_{3,i}^\dagger D_{3,i}+E_{1,i}^\dagger E_{1,i}+F_{2,i}^\dagger F_{2,i}\right)+\frac{\varepsilon_2}{2}\sum_{i}\left( A_{1,i}^\dagger A_{1,i}+A_{2,i}^\dagger A_{2,i}\right.\nonumber\\
 &\left.+ B_{2,i}^\dagger B_{2,i}+B_{3,i}^\dagger B_{3,i}+ C_{1,i}^\dagger C_{1,i}+C_{3,i}^\dagger C_{3,i}+D_{1,i}^\dagger D_{1,i}+D_{2,i}^\dagger D_{2,i}+E_{2,i}^\dagger E_{2,i}+E_{3,i}^\dagger E_{3,i}+F_{1,i}^\dagger F_{1,i}+F_{3,i}^\dagger F_{3,i}\right)\nonumber\\
&+\frac{\varepsilon_3}{2}\sum_{i} \left[(a^1_{1,i})^\dagger a^1_{1,i}+(a^2_{3,i})^\dagger a^2_{3,i}+(b^1_{1,i})^\dagger b^1_{1,i}+(b^2_{2,i})^\dagger b^2_{2,i}+(c^1_{3,i})^\dagger c^1_{3,i}+(c^2_{2,i})^\dagger c^2_{2,i}+(d^1_{3,i})^\dagger d^1_{3,i}+(d^2_{1,i})^\dagger d^2_{1,i}\right.\nonumber\\
&\left.+(e^1_{2,i})^\dagger e^1_{2,i}+(e^2_{1,i})^\dagger e^2_{1,i}+(f^1_{2,i})^\dagger f^1_{2,i}+(f^2_{3,i})^\dagger f^2_{3,i}\right]+\frac{\varepsilon_4}{2}\sum_{i}\left[ (a^1_{2,i})^\dagger a^1_{2,i}+(a^2_{1,i})^\dagger a^2_{1,i}+(b^1_{2,i})^\dagger b^1_{2,i}\right.\nonumber\\
&\left.+(b^2_{3,i})^\dagger b^2_{3,i}+(c^1_{1,i})^\dagger c^1_{1,i}+(c^2_{3,i})^\dagger c^2_{3,i}+(d^1_{1,i})^\dagger d^1_{1,i}+(d^2_{2,i})^\dagger d^2_{2,i}+(e^1_{3,i})^\dagger e^1_{3,i}+(e^2_{2,i})^\dagger e^2_{2,i}+(f^1_{3,i})^\dagger f^1_{3,i}+(f^2_{1,i})^\dagger f^2_{1,i}\right]\nonumber\\
&+\frac{\varepsilon_5}{2}\sum_{i} \left[(a^1_{3,i})^\dagger a^1_{3,i}+(a^2_{2,i})^\dagger a^2_{2,i}+(b^1_{3,i})^\dagger b^1_{3,i}+(b^2_{1,i})^\dagger b^2_{1,i}+(c^1_{2,i})^\dagger c^1_{2,i}+(c^2_{1,i})^\dagger c^2_{1,i}+(d^1_{2,i})^\dagger d^1_{2,i}+(d^2_{3,i})^\dagger d^2_{3,i}\right.\nonumber\\
&\left.+(e^1_{1,i})^\dagger e^1_{1,i}+(e^2_{3,i})^\dagger e^2_{3,i}+(f^1_{1,i})^\dagger f^1_{1,i}+(f^2_{2,i})^\dagger f^2_{2,i}\right]+V_5\sum_{i}\left[A_{3,i}^\dagger(A_{1,i}+A_{2,i})+B_{1,i}^\dagger(B_{2,i}+B_{3,i})+C_{2,i}^\dagger(C_{1,i}\right.\nonumber\\
&\left.+C_{3,i})+D_{3,i}^\dagger(D_{1,i}+D_{2,i})+E_{1,i}^\dagger(E_{2,i}+E_{3,i})+F_{2,i}^\dagger(F_{1,i}+F_{3,i})\right]+V_6\sum_i\left[ A_{1,i}^\dagger A_{2,i}+B_{2,i}^\dagger B_{3,i}+C_{1,i}^\dagger C_{3,i}\right.\nonumber\\
&\left.+ D_{1,i}^\dagger D_{2,i}+E_{2,i}^\dagger E_{3,i}+F_{1,i}^\dagger F_{3,i}\right]+V_7\sum_i \left[(a^1_{2,i})^\dagger a^1_{3,i}+(a^2_{1,i})^\dagger a^2_{2,i}+(b^1_{2,i})^\dagger b^1_{3,i}+(b^2_{3,i})^\dagger b^2_{1,i}+(c^1_{1,i})^\dagger c^1_{2,i}\right.\nonumber\\
&\left.+(c^2_{3,i})^\dagger c^2_{1,i}+(d^1_{1,i})^\dagger d^1_{2,i}+(d^2_{2,i})^\dagger d^2_{3,i}+(e^1_{1,i})^\dagger e^1_{3,i}+(e^2_{2,i})^\dagger e^2_{3,i}+(f^1_{3,i})^\dagger f^1_{1,i}+(f^2_{1,i})^\dagger f^2_{2,i}\right]+V_8\sum_i \left[(a^1_{1,i})^\dagger a^1_{3,i}
\right.\nonumber\\
&\left.+(a^2_{3,i})^\dagger a^2_{2,i}+(b^1_{1,i})^\dagger b^1_{3,i}+(b^2_{2,i})^\dagger b^2_{1,i}+(c^1_{3,i})^\dagger c^1_{2,i}+(c^2_{2,i})^\dagger c^2_{1,i}+(d^1_{3,i})^\dagger d^1_{2,i}+(d^2_{1,i})^\dagger d^2_{3,i}+(e^1_{1,i})^\dagger e^1_{2,i}+(e^2_{1,i})^\dagger e^2_{3,i}\right.\nonumber\\
&\left.+(f^1_{2,i})^\dagger f^1_{1,i}+(f^2_{3,i})^\dagger f^2_{2,i}\right]+V_9\sum_i \left[(a^1_{1,i})^\dagger a^1_{2,i}+(a^2_{3,i})^\dagger a^2_{1,i}+(b^1_{1,i})^\dagger b^1_{2,i}+(b^2_{2,i})^\dagger b^2_{3,i}+(c^1_{3,i})^\dagger c^1_{1,i}+(c^2_{2,i})^\dagger c^2_{3,i}\right.\nonumber\\
&\left.+(d^1_{3,i})^\dagger d^1_{1,i}+(d^2_{1,i})^\dagger d^2_{2,i}+(e^1_{3,i})^\dagger e^1_{2,i}+(e^2_{1,i})^\dagger e^2_{2,i}+(f^1_{2,i})^\dagger f^1_{3,i}+(f^2_{3,i})^\dagger f^2_{1,i}\right]+h.c.
\end{align}
In the absence of SOC, we actually do not need to consider the $\sigma$-orbitals since they correspond to states away from the Fermi level, and we are interested in the physics around the Fermi energy.
\subsection{Tight-binding model with spin-orbit coupling}
Upon including SOC, the Hamiltonian changes into
\begin{align}
H_{\rm full}&=H_0+H^{z,\sigma}_{SOC}+(H^{z,\sigma}_{SOC})^\dagger,
\end{align}
where $H^{z,\sigma}_{SOC}$ accounts for the hopping from $p_z$ orbitals to $\sigma$-orbitals due to the SOC. The latter can be decomposed as
\begin{align}\label{eqn4}
H^{z,\sigma}_{SOC}&=H^{z,\sigma}_L+H^{z,\sigma}_E.
\end{align}
The first term, $H^{z,\sigma}_L$, originates from relativistic corrections to the Schr\"odinger equation described by the Hamiltonian $H_{L}=-f(r){\bm\sigma}\cdot {\bf L}$, where $f$ is a function related to the effective short-ranged nuclear electrostatic potential, ${\bm \sigma}$ is a vector of Pauli matrices, and ${\bf L}$ is the orbital angular momentum, see also Table~\ref{tablematrix} for the matrix elements of this Hamiltonian in the basis we use here. This term will give rise to the intrinsic SOC. The second term, $H^{z,\sigma}_E$, results from an applied electric field ${E}$ perpendicular to the lattice plane that breaks the reflection symmetry through it, and this term is associated with the Rashba SOC. As a result, we obtain
\begin{align}
H^{z,\sigma}_{E}&=\sum_{j,\alpha=1,2}\xi_{sp\alpha}p^\dagger_{z,j\alpha}s_{j\alpha},\nonumber\\
H^{z,\sigma}_{L}&=\sum_{j,\alpha=1,2}\xi_{p\alpha}p^\dagger_{z,j\alpha}\left(-i\sigma_y p_{x,j\alpha}+i\sigma_x p_{y,j\alpha}\right).\nonumber
\end{align}
Here, $p_{z,i}^\dagger$ creates an electron in a $p_z$ orbital at position $i$, and analogous notation is used for the $p_x$, $p_y$, and $s$ orbitals, while $\alpha=1(2)$ corresponds to the sites at vertex (edge). The variables $\xi_{sp\alpha}$ and $\xi_{p\alpha}$ denote coupling constants, the values of which should be determined by first-principle calculations.
\begin{table}[t]
\centering
\begin{tabular}{ l || c | c | c  }
   ${\bm\sigma}\cdot {\bm L}$& $p_x$ & $p_y$ & $s$ \\
  \hline\hline
$p_z$& $-i\sigma_y$  & $i\sigma_x$ & $0$
\end{tabular}
\caption{Matrix elements for ${\bm\sigma}\cdot{\bf L}$. Note that the Pauli matrices act in spin space.}
\label{tablematrix}
\end{table}
In our studies, we explicitly make use of the hybrid orbitals, therefore we need to rewrite the above microscopic Hamiltonians in terms of these. In Table~\ref{basis} we provide the convention we used for the change of basis, which leads to
\begin{table}[b]
\resizebox{\linewidth}{!}{%
\begin{tabular}{c||c|c|c|c|c|c|c|c|c|c|c|c|c|c|c|c|c|c}
    &\multicolumn{6}{c|}{$sp^2$}&\multicolumn{6}{c|}{$sp$}&\multicolumn{6}{|c}{$p$}\\\hline
&\includegraphics[scale=0.8]{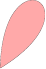} &\includegraphics[scale=0.8]{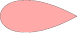}&\includegraphics[scale=0.8]{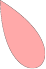}&\includegraphics[scale=0.8]{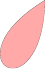}&\includegraphics[scale=0.8]{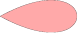}&\includegraphics[scale=0.8]{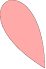}&\includegraphics[scale=0.8]{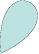}&\includegraphics[scale=0.8]{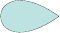}&\includegraphics[scale=0.8]{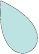}&\includegraphics[scale=0.8]{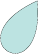} &\includegraphics[scale=0.8]{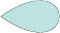}&\includegraphics[scale=0.8]{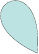}&\includegraphics[scale=0.8]{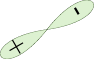}&\includegraphics[scale=0.8]{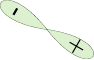}&\includegraphics[scale=0.8]{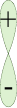}&\includegraphics[scale=0.8]{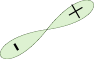}&\includegraphics[scale=0.8]{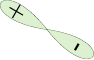}&\includegraphics[scale=0.8]{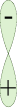}\\\hline
$ s$&$1/\sqrt{3}$&$1/\sqrt{3}$&$1/\sqrt{3}$&$1/\sqrt{3}$&$1/\sqrt{3}$&$1/\sqrt{3}$&$1/\sqrt{2}$&$1/\sqrt{2}$&$1/\sqrt{2}$&$1/\sqrt{2}$&$1/\sqrt{2}$&$1/\sqrt{2}$&0&0&0&0&0&0\\
$ p_x$&$1/\sqrt{6}$&$-\sqrt{2/3}$&$1/\sqrt{6}$&$-1/\sqrt{6}$&$\sqrt{2/3}$&$-1/\sqrt{6}$&$1/\sqrt{8}$&$-1/\sqrt{2}$&$1/\sqrt{8}$&$-1/\sqrt{8}$&$1/\sqrt{2}$&$-1/\sqrt{8}$&$-\sqrt{3}/2$&$\sqrt{3}/2$&$0$&$\sqrt{3}/2$&$-\sqrt{3}/2$&$0$\\
$ p_y$&$1/\sqrt{2}$&$0$&$-1/\sqrt{2}$&$-1/\sqrt{2}$&$0$&$1/\sqrt{2}$&$\sqrt{3/8}$&$0$&$-\sqrt{3/8}$&$-\sqrt{3/8}$&$0$&$\sqrt{3/8}$&$-1/2$&$-1/2$&$1$&$1/2$&$1/2$&$-1$
\end{tabular}}
\caption{Hybridized $sp^2$, $sp$, and $p$ orbitals in terms of the atomic $s$, $p_x$, and $p_y$ orbitals. For example, orbital $A_1$ in Fig.\ \ref{figa}(b) reads $|A_1\rangle=\frac{1}{\sqrt{3}}|s\rangle-\frac{1}{\sqrt{6}}|p_x\rangle+\frac{1}{\sqrt{2}}|p_y\rangle$.}
\label{basis}
\end{table}
\begin{align}
H^{z,\sigma}_{E}&=\xi_{sp1}3^{-1/2}\sum_i\left[ A_i^\dagger(A_{1,i}+A_{2,i}+A_{3,i})+B_i^\dagger(B_{1,i}+B_{2,i}+B_{3,i})+C_i^\dagger(C_{1,i}+C_{2,i}+C_{3,i})+D_i^\dagger(D_{1,i}+D_{2,i}\right.\nonumber\\
&\left.+D_{3,i})+E_i^\dagger(E_{1,i}+E_{2,i}+E_{3,i})+F_i^\dagger(F_{1,i}+F_{2,i}+F_{3,i})\right]+\xi_{sp2}2^{-1/2}\sum_i\left[ (a^1_{i})^\dagger(a^1_{2,i}+a^1_{1,i})+(a^2_{i})^\dagger(a^2_{1,i}+a^2_{3,i})\right.\nonumber\\
&+(b^1_{i})^\dagger(b^1_{2,i}+b^1_{1,i})+(b^2_{i})^\dagger(b^2_{2,i}+b^2_{3,i})+(c^1_{i})^\dagger(c^1_{1,i}+c^1_{3,i})+(c^2_{i})^\dagger(c^2_{2,i}+c^2_{3,i})+(d^1_{i})^\dagger(d^1_{1,i}+d^1_{3,i})\nonumber\\
&\left.+(d^2_{i})^\dagger(d^2_{1,i}+d^2_{2,i})+(e^1_{i})^\dagger(e^1_{3,i}+e^1_{2,i})+(e^2_{i})^\dagger(e^2_{1,i}+e^2_{2,i})+(f^1_{i})^\dagger(f^1_{2,i}+f^1_{3,i})+(f^2_{i})^\dagger(f^2_{1,i}+f^2_{3,i})\right],
\end{align}
and
\begin{align}
H_{L}^{z,\sigma}&=i\xi_{p1}\sum_{i}\left[A^\dagger_{i}( 2^{-1/2}\sigma_x+6^{-1/2}\sigma_y)A_{1,i}+A^\dagger_{i}(- (2/3)^{1/2}\sigma_y)A_{2,i}+A^\dagger_{i}(- 2^{-1/2}\sigma_x+6^{-1/2}\sigma_y)A_{3,i}\right.\nonumber\\
&+B^\dagger_{i}(-2^{-1/2}\sigma_x-6^{-1/2}\sigma_y)B_{1,i}+B^\dagger_{i}( (2/3)^{1/2}\sigma_y)B_{2,i}+B^\dagger_{i}( 2^{-1/2}\sigma_x-6^{-1/2}\sigma_y)B_{3,i}\nonumber\\
&+C^\dagger_{i}( 2^{-1/2}\sigma_x+6^{-1/2}\sigma_y)C_{1,i}+C^\dagger_{i}(-(2/3)^{1/2}\sigma_y)C_{2,i}+C^\dagger_{i}(-2^{-1/2}\sigma_x+6^{-1/2}\sigma_y)C_{3,i}\nonumber\\
&+D^\dagger_{i}(- 2^{-1/2}\sigma_x-6^{-1/2}\sigma_y)D_{1,i}+D^\dagger_{i}( (2/3)^{1/2}\sigma_y)D_{2,i}+D^\dagger_{i}( 2^{-1/2}\sigma_x-6^{-1/2}\sigma_y)D_{3,i}\nonumber\\
&+E^\dagger_{i}( 2^{-1/2}\sigma_x+6^{-1/2}\sigma_y)E_{1,i}+E^\dagger_{i}(-(2/3)^{1/2}\sigma_y)E_{2,i}+E^\dagger_{i}(-2^{-1/2}\sigma_x+6^{-1/2}\sigma_y)E_{3,i}\nonumber\\
&\left.+F^\dagger_{i}(- 2^{-1/2}\sigma_x-6^{-1/2}\sigma_y)F_{1,i}+F^\dagger_{i}( (2/3)^{1/2}\sigma_y)F_{2,i}+F^\dagger_{i}( 2^{-1/2}\sigma_x-6^{-1/2}\sigma_y)F_{3,i}\right]\nonumber\\
&+i\xi_{p2}\sum_i\left[(a^1_{i})^\dagger(-2^{-1/2}\sigma_y)a^1_{1,i}+(a^1_{i})^\dagger(2^{-1/2}\sigma_y)a^1_{2,i}+(a^1_{i})^\dagger(\sigma_x)a^1_{3,i}+(a^2_{i})^\dagger2^{-1/2}(-\sqrt{3}\sigma_x/2-\sigma_y/2)a^2_{1,i}\right.\nonumber\\
&+(a^2_{i})^\dagger(\sigma_x/2-\sqrt{3}\sigma_y/2)a^2_{2,i}+(a^2_{i})^\dagger2^{-1/2}(\sqrt{3}\sigma_x/2+\sigma_y/2)a^2_{3,i}+(b^1_{i})^\dagger2^{-1/2}\sigma_y b^1_{1,i}-(b^1_{i})^\dagger2^{-1/2}\sigma_y b^1_{2,i}\nonumber\\
&+(b^1_{i})^\dagger\sigma_x b^1_{3,i}+(b^2_{i})^\dagger (\sigma_x/2+\sqrt{3}\sigma_y/2)b^2_{1,i}+(b^2_{i})^\dagger 2^{-1/2}(\sqrt{3}\sigma_x/2-\sigma_y/2)b^2_{2,i}-(b^2_{i})^\dagger 2^{-1/2}(\sqrt{3}\sigma_x/2-\sigma_y/2)b^2_{3,i}\nonumber\\
&+(c^2_{i})^\dagger(\sigma_x/2+\sqrt{3}\sigma_y/2)c^2_{1,i}-(c^2_{i})^\dagger 2^{-1/2}(\sqrt{3}\sigma_x/2-\sigma_y/2)c^2_{2,i}+(c^2_{i})^\dagger 2^{-1/2}(\sqrt{3}\sigma_x/2-\sigma_y/2)c^2_{3,i}\nonumber\\
&+(c^1_{i})^\dagger2^{-1/2}(-\sqrt{3}\sigma_x/2-\sigma_y/2)c^1_{1,i}-(c^1_{i})^\dagger(\sigma_x/2-\sqrt{3}\sigma_y/2)c^1_{2,i}+(c^1_{i})^\dagger2^{-1/2}(\sqrt{3}\sigma_x/2+\sigma_y/2)c^1_{3,i}\nonumber\\
&-(d^1_{i})^\dagger2^{-1/2}(-\sqrt{3}\sigma_x/2-\sigma_y/2)d^1_{1,i}-(d^1_{i})^\dagger(\sigma_x/2-\sqrt{3}\sigma_y/2)d^1_{2,i}-(d^1_{i})^\dagger2^{-1/2}(\sqrt{3}\sigma_x/2+\sigma_y/2)d^1_{3,i}\nonumber\\
&-(d^2_{i})^\dagger(-2^{-1/2}\sigma_y)d^2_{1,i}-(d^2_{i})^\dagger(2^{-1/2}\sigma_y)d^2_{2,i}-(d^2_{i})^\dagger(\sigma_x)d^2_{3,i}\nonumber\\
&-(e^1_{i})^\dagger (\sigma_x/2+\sqrt{3}\sigma_y/2)e^1_{1,i}-(e^1_{i})^\dagger 2^{-1/2}(\sqrt{3}\sigma_x/2-\sigma_y/2)e^1_{2,i}+(e^1_{i})^\dagger 2^{-1/2}(\sqrt{3}\sigma_x/2-\sigma_y/2)e^1_{3,i}\nonumber\\
&-(e^2_{i})^\dagger2^{-1/2}\sigma_y e^2_{1,i}+(e^2_{i})^\dagger2^{-1/2}\sigma_y e^2_{2,i}-(e^2_{i})^\dagger\sigma_x e^2_{3,i}\nonumber\\
&-(f^1_{i})^\dagger (\sigma_x/2+\sqrt{3}\sigma_y/2)f^1_{1,i}+(f^1_{i})^\dagger 2^{-1/2}(\sqrt{3}\sigma_x/2-\sigma_y/2)f^1_{2,i}-(f^1_{i})^\dagger 2^{-1/2}(\sqrt{3}\sigma_x/2-\sigma_y/2)f^1_{3,i}\nonumber\\
&\left.(f^2_{i})^\dagger2^{-1/2}(\sqrt{3}\sigma_x/2+\sigma_y/2)f^2_{1,i}+(f^2_{i})^\dagger(\sigma_x/2-\sqrt{3}\sigma_y/2)f^2_{2,i}-(f^2_{i})^\dagger2^{-1/2}(\sqrt{3}\sigma_x/2+\sigma_y/2)f^2_{3,i}\right].
\end{align}
Since we are ultimately interested in the physics around the Fermi level, we integrate out the $\sigma$-orbitals to obtain an effective Hamiltonian solely in terms of the 18 $p_z$ orbitals, see Section~\ref{efff}. The procedure outlined in Sec. III then leads to the following effective Hamiltonian:
\begin{align}
H^{\rm eff}_{z,18}&=S^{-1/2}\left[H_0^z-H^{z,\sigma}_{SOC}(H^\sigma_0)^{-1}(H^{z,\sigma}_{SOC})^\dagger\right]S^{-1/2},
\end{align}
where $S=\mathbb{I}+H^{z,\sigma}_{SOC}(H^\sigma_0)^{-2}(H^{z,\sigma}_{SOC})^\dagger$. We treat the term $S^{-1/2}$ as a power series, i.e. \begin{align}
S^{-1/2}&=\mathbb{I}-H^{z,\sigma}_{SOC}(H^\sigma_0)^{-2}(H^{z,\sigma}_{SOC})^\dagger/2+\ldots
\end{align}
As a result, we obtain
\begin{align}
H^{\rm eff}_{z,18}&=S^{-1/2}(H^z_0-H^{z,\sigma}_{SOC}(H^\sigma_0)^{-1}(H^{z,\sigma}_{SOC})^\dagger)S^{-1/2}\nonumber\\
&=(\mathbb{I}-H^{z,\sigma}_{SOC}(H^\sigma_0)^{-2}(H^{z,\sigma}_{SOC})^\dagger/2)(H^z_0-H^{z,\sigma}_{SOC}(H^\sigma_0)^{-1}(H^{z,\sigma}_{SOC})^\dagger)(\mathbb{I}-H^{z,\sigma}_{SOC}(H^\sigma_0)^{-2}(H^{z,\sigma}_{SOC})^\dagger/2)+\ldots\nonumber\\
&=(\mathbb{I}-H^{z,\sigma}_{SOC}(H^\sigma_0)^{-2}(H^{z,\sigma}_{SOC})^\dagger/2)H^z_0(\mathbb{I}-H^{z,\sigma}_{SOC}(H^\sigma_0)^{-2}(H^{z,\sigma}_{SOC})^\dagger/2)\nonumber\\
&-(\mathbb{I}-H^{z,\sigma}_{SOC}(H^\sigma_0)^{-2}(H^{z,\sigma}_{SOC})^\dagger/2)H^{z,\sigma}_{SOC}(H^\sigma_0)^{-1}(H^{z,\sigma}_{SOC})^\dagger(\mathbb{I}-H^{z,\sigma}_{SOC}(H^\sigma_0)^{-2}(H^{z,\sigma}_{SOC})^\dagger/2)+\ldots
\end{align}
In the last line, we have split the Hamiltonian in two parts. One can easily see that if one neglects the last term, the Hamiltonian gets altered, but neither gaps will open nor will the positions of the Dirac cones shift. This is a simple consequence of the fact that $\det{(S^{-1/2}H_0^zS^{-1/2})}=\det{(H_0^z)}/\det{(S)}$. As a result, if we simply assume $S=\mathbb{I}$ we do not miss either any opening of a bandgap or a shift in the position of a Dirac cone, which we are ultimately interested in here.
With respect to the second term, we would like to point out that $H^{z,\sigma}_{SOC}(H^\sigma_0)^{-1}(H^{z,\sigma}_{SOC})^\dagger$ and $H^{z,\sigma}_{SOC}(H^\sigma_0)^{-2}(H^{z,\sigma}_{SOC})^\dagger$ are both proportional to $\xi^2$, with $\xi\in\{\xi_{sp1},\xi_{sp2},\xi_{p1},\xi_{p2}\}$. Therefore, using that $S^{-1/2}\approx \mathbb{I}-H^{z,\sigma}_{SOC}(H^\sigma_0)^{-2}(H^{z,\sigma}_{SOC})^\dagger/2$, we find a contribution proportional to $\xi^4$, which can be neglected since all $\xi$'s are very small compared to the other hopping parameters. Hence, we may simply set $S=\mathbb{I}$, and approximate the effective Hamiltonian by
\begin{align}\label{13}
H^{\rm eff}_{z,18}&\approx H_0^z-H^{z,\sigma}_{SOC}(H_0^\sigma)^{-1}(H^{z,\sigma}_{SOC})^\dagger.
\end{align}
Having argued that we may set $S=\mathbb{I}$, we now show how to deal with $(H_0^\sigma)^{-1}$. Because the hybrid orbitals are mainly composed of $p_x$ and $p_y$ orbitals, which have an almost vanishing onsite energy [S1], it holds that $\rVert H_{\sigma,{\rm onsite}}\lVert<\rVert H_{\sigma,{\rm NN}}\lVert$. Therefore, we can write $(H_0^{\sigma})^{-1}\approx H_{\sigma,{\rm NN}}^{-1}-H_{\sigma,{\rm NN}}^{-1}H_{\sigma,{\rm onsite}}H_{\sigma,{\rm NN}}^{-1}$, and
\begin{align}
\left(H_{\sigma,\rm{NN}}^{-1}\right)_{ij}=\left\{
	\begin{array}{ll}
		1/\left(H_{\sigma,\rm{NN}}\right)^*_{ij}  & \mbox{if } (H_{\sigma,\rm{NN}})_{ij} \neq 0 \\
		0 & \mbox{if } (H_{\sigma,\rm{NN}})_{ij} = 0.
	\end{array}\right.
\end{align}
One may easily transform this expression to real space. The effective SOC Hamiltonian is then given by the second term on the RHS of Eq.~(\ref{13}),
\begin{align}\label{repl2}
H_{\rm{SOC,eff}}&=H^{z,\sigma}_{SOC}\left(H_{\sigma,\rm{NN}}^{-1}-H_{\sigma,\rm{NN}}^{-1}H_{\sigma,\rm{onsite}}H_{\sigma,\rm{NN}}^{-1}\right)(H_{SOC}^{z,\sigma})^\dagger.
\end{align}
Replacing the Hamiltonians $H_{\sigma,{\rm NN}}$, $H_{\sigma,\rm{onsite}}$, and $H^{z,\sigma}_{SOC}$ given by Eqs.~(\ref{eqn2}),~(\ref{eqn3}), and~(\ref{eqn4}), respectively into Eq.~(\ref{repl2}) we obtain the effective SOC Hamiltonian
\begin{align}\label{eq8}
H_{\rm{SOC, eff}}=H_{\rm{R}}+H_{\rm{I}}+H_{\rm{rest}},
\end{align}
where
\begin{align}
H_{\rm{R}}&=i\sum_{\langle i,j\rangle}\lambda_{R,ij}p_{z,i}^\dagger\left({\bm\sigma}\times{\bf\hat{d}}_{ij}\right)\cdot{\bf\hat{z}}p_{z,j},\\
H_{\rm{I}}&=i\sum_{\langle\langle i,j\rangle\rangle}\lambda_{I,ij}v_{ij}p_{z,i}^\dagger \sigma_z p_{z,j}.
\end{align}
Here, $H_{R}$ and $H_I$ denote, respectively, the Rashba and intrinsic SOC. $H_{\rm{rest}}$ describes next-nearest neighbor (NNN) corrections to the Rashba SOC (which can be neglected), and some very small spin independent and thus irrelevant NN and NNN hoppings and on-site energies. In both expressions we have included position-dependent coupling constants $\lambda_{R,ij}$ and $\lambda_{I,ij}$, ${\bf\hat{d}}_{ij}$ is the unit vector pointing from site $i$ to $j$, and $v_{ij}=+(-)$ if the hopping is (anti)-clockwise, and zero if it is along the acetylene bond. There are three coupling parameters for the Rashba SOC ($\lambda_{R,i}$), with $i=1,2,3$, corresponding to vertex-vertex,  vertex-edge, and edge-edge hoppings, respectively. Furthermore, there are two coupling parameters for the intrinsic SOC ($\lambda_{I,j}$), with $j=2,3$ corresponding to vertex-edge and edge-edge hoppings, respectively. In Table \ref{table1} we give the expressions for these coupling parameters in terms of the $\sigma$-hopping parameters, as obtained from Eq.~(\ref{eq8}). It turns out that in $\beta$-graphyne there is an additional term, which results from the absence of  mirror symmetry through the acetylene bond, given by
\begin{align}
H_{\beta,4}&=i\lambda_{I,4}\sum_{\langle\langle i,j\rangle\rangle}w_{ij}p^\dagger_{z,i}\sigma_z p_{z,j},
\end{align}
where $w_{ij}=+(-)$ if the hopping is along the acetylene bond going (anti-)clockwise with respect to the center of the unit-cell.
\begin{table}[t]

\centering
\begin{tabular}{ l || c  }
   $\sigma$& microscopic parameters\\
  \hline\hline
  $\lambda_{R,1}$ & $2\sqrt{2}\xi_{sp2}\xi_{p2}/(3V_1)$ \\
  $\lambda_{R,2}$ & $(\sqrt{2}\xi_{sp1}\xi_{p2}+\xi_{sp2}\xi_{p1})/(\sqrt{6}V_2)$\\
  $\lambda_{R,3}$ & $\xi_{1sp}\xi_{p1}/V_3$ \\
  $\lambda_{I,2}$  & $V_{5}\xi_{p2}\xi_{p1}/(2V_1 V_2)$\\
  $\lambda_{I,3}$ & $\sqrt{3}V_6\xi_{p1}^2/(4V_2^2)$ \\
  $\lambda_{I,4}$ & $\sqrt{2}\xi_{p1}\xi_{p2}V_7/(\sqrt{3}V_2 V_4)$ \\
\end{tabular}
\caption{SOC parameters for $\beta$-graphyne arising from the $\sigma$-orbitals.}
\label{table1}
\end{table}\\

The next step is  to integrate out the $12$ $p_z$-orbitals located at the edges, which yields
\begin{align}\label{effsoc}
H^{\rm{eff}}_{z,6}&=S^{-1/2}\left(H_{vv}-H_{ve}H_{ee}^{-1}H_{ve}^\dagger\right)S^{-1/2},
\end{align}
where $H_{vv}$($H_{ee}$) describes the $p_z$ orbitals at the vertices (edges), and $H_{ve}$ couples the two sets of orbitals. To account for the SOC, we write each matrix as the sum of a spin-independent part, denoted by a tilde, and a part describing the SOC, denoted by the subscript SOC, e.g., $H_{vv}=\tilde{H}_{vv}+H_{vv,SOC}$. Because the SOC parameters are very small compared to the other hopping energies, we may expand Eq.~(\ref{effsoc}) up to first order in $\lambda_{R,i}$ and $\lambda_{I,j}$. We then obtain
\begin{align}
H^{\rm{eff}}_{z,6}&\approx\tilde{H}^{\rm{eff}}_{z,6}+H_{SOC},
\end{align}
where $H_{SOC}=H_{1,SOC}+H_{2,SOC}+H_{3,SOC}+H_{4,SOC}$, with
\begin{align}\label{eq2}
H_{1,SOC}&=-\frac{1}{2}\tilde{S}^{-3/2}S_{SOC}\left(\tilde{H}_{vv}-\tilde{H}_{ve}\tilde{H}_{ee}^{-1}\tilde{H}_{ve}^\dagger\right)\tilde{S}^{-1/2}+h.c.,\\
H_{2,SOC}&=\tilde{S}^{-1/2}H_{vv,SOC}\tilde{S}^{-1/2},\\
H_{3,SOC}&=-\tilde{S}^{-1/2}H_{ve,SOC}\tilde{H}_{ee}^{-1}\tilde{H}_{ve}^\dagger\tilde{S}^{-1/2}+h.c.,\\
H_{4,SOC}&=\tilde{S}^{-1/2}\tilde{H}_{ve}\tilde{H}_{ee}^{-1}H_{ee,SOC}\tilde{H}_{ee}^{-1}\tilde{H}_{ve}^\dagger\tilde{S}^{-1/2}.\label{eq3}
\end{align}
It can already be seen that $H_{1,SOC}$ is proportional to $\tilde{S}^{-2}$ and therefore very small compared to the other terms. As a result, we neglect this term. As compared to graphene, we get a richer SOC since now we have to distinguish between the inter and intra-unit cell SOC. We refer to the inter-(intra-)unit cell SOC as external (internal) SOC. The form of the SOC Hamiltonians is however unchanged as compared to graphene,
 \begin{align}\label{intext}
 H_{\rm R}^{\rm eff}&=i \sum_{a,\langle i,j\rangle}\lambda_{R,a}\, p_{z,i}^\dagger \, ({\bm \sigma}\times \hat{{\bf d}}_{ij})\cdot\hat{{\bf z}}\, p_{z,j},\\
H_{\rm I}^{\rm eff}&=i \sum_{a,\langle\langle i,j\rangle\rangle}\,\lambda_{L,a}\,v_{ij}\,p_{z,i}^\dagger\,\sigma_z\, p_{z,j}.\label{intext2}
\end{align}
Here, the index $a={\rm int}({\rm ext})$ refers to the SOC's effectively described by the intra-(inter-)unit-cell hoppings, with the summations also taken correspondingly. In Table~\ref{table8}, we have listed the expression for the SOC hopping parameters. Combining Tables~\ref{table1} and \ref{table8}, we obtain the effective parameters displayed in Table I of the main text.
\begin{table*}[t]
\centering
\begin{tabular}{ l || c  }
  SOC &SOC parameters for six-site model\\
  \hline\hline

  $\lambda_{\rm I,ext}$  & $-\lambda_{I,2}t_2 t_3 /(t_3^2+2t_2^2)$\\

  $\lambda_{\rm I,int}$ & $\lambda_{I,3}t_2^2/(t_3^2+2t_2^2)$\\

  $\lambda_{\rm R,ext}$ & $\lambda_{R,1}t_3^2/(t_3^2+2t_2^2)$\\

  $\lambda_{\rm R,int}$ & $-2\lambda_{R,2}t_2t_3/(t_3^2+2t_2^2)-\lambda_{R,3}t_2^2/(t_3^2+2t_2^2)$\\

\end{tabular}
\caption{SOC parameters for the effective TB models.}
\label{table8}
\end{table*}\\

Inspection of Table~\ref{table1} clearly shows that the coupling parameters have a particular form. All the Rashba coupling parameters are proportional to $\xi_{sp}\xi_{p}/V_i$, where $V_i$ is one of the NN hopping parameters. This is expected, since the Rashba terms arise from
\begin{align}\label{27}
H^{z,\sigma}_E H_{\sigma,\rm{NN}}^{-1}(H^{z,\sigma}_L)^\dagger+h.c.,
\end{align}
where the matrix $H^{z,\sigma}_E$ contributes a factor $\xi_{sp}$, $H_{\rm{NN}}$ yields a factor $V_i$, and $H^{z,\sigma}_L$ a factor $\xi_p$. Note that $V_4$ does not appear in Table~\ref{table1}; this can be understood from the matrix structure of Eq.~(\ref{27}). Terms from right to left in this expression correspond to {\it(i)} the onsite hopping from a $p_z$ orbital to a hybrid orbital, {\it (ii)} subsequent hopping between two NN hybrid orbitals, and {\it(iii)} the onsite hopping from a $sp$ or $sp^2$ hybrid orbital to a $p_z$ orbital due to the electric field. However, $V_4$ is responsible for the hopping between two NN $p$ orbitals and thus does not contribute to this process. Notice that $\varepsilon_1-\varepsilon_5$ and $V_5-V_9$ do not appear here because they correspond to onsite energies and hoppings, repectively.\\

The intrinsic SOC arises from
\begin{align}\label{eq1}
H^{z,\sigma}_L H_{\sigma,\rm{NN}}^{-1}H_{\sigma,\rm{onsite}}H_{\sigma,\rm{NN}}^{-1}(H^{z,\sigma}_L)^\dagger+h.c.,
\end{align}
hence the coupling parameters are all of the form $\xi_p^2 A B^{-2}$, where $A\in\{V_5,\ldots,V_9,\varepsilon_1,\ldots,\varepsilon_5\}$ comes from $H_{\sigma,\rm{onsite}}$ and $B\in\{V_1,\ldots,V_4\}$ comes from $H_{\sigma,\rm{NN}}$. However, Table~\ref{table1} clearly shows that actually none of these parameters is proportional to an onsite energy $\varepsilon_i$. The reason for this is that if we read Eq.~(\ref{eq1}) from right to left, we find that {\it (i)} the first matrix leads to the hopping from a $p_z$ orbital to one of the hybrid orbitals due to the SOC, {\it (ii)} then $H_{\sigma,\rm{NN}}^{-1}$ leads to the hopping to a NN hybrid orbital,{\it (iii,a)} the term $H_{\sigma,\rm{onsite}}$ can lead to the onsite hopping to another hybrid orbital; this contributes a factor $V_j$ with $j=5,\ldots,9$,{\it (iii,b)} or it can simply stay on the same orbital which would contribute a factor $\varepsilon_i$. However, this last scenario leads then subsequently to the hopping to the hybrid orbital  where one started, therefore there are no terms proportional to $\varepsilon_i$. The parameter $V_8$  is also absent in Table~\ref{table1}. This can be seen from studying the hopping process proportional to $V_8$, starting from orbital $a_1$,  we then find that
\begin{enumerate}
\item $(H^{z,\sigma}_{L})^\dagger$ contributes a factor proportional to $\sigma_y$, and leads to the hopping to state $a^1_1$
\item $H_{\sigma,\rm{NN}}^{-1}$ contributes a factor $V_3^{-1}$, and leads to the hopping to state $b^1_1$
\item $H_{\sigma,\rm{onsite}}$ contributes a factor $V_8$, and leads to the hopping to state $b^1_3$
\item $H_{\sigma,\rm{NN}}^{-1}$ contributes a factor $V_3^{-1}$, and leads to the hopping to state $a^1_1$
\item $(H^{z,\sigma}_{L})^\dagger$ contributes a factor proportional to $\sigma_y$, and leads to the hopping to state $a_1$.
 \end{enumerate}
 Hence, the term proportional to $V_8$ does not lead to a NNN hopping process,  but gives rise to an onsite energy, which can be neglected. With regards to $V_9$, we would like to point out that this term does actually lead to a NNN hopping in Eq.~(\ref{eq1}), however this is a spin-independent hopping. This can be seen from the fact that from right to left again, for this process the hybrid orbital to which the $p_z$ orbital hops points in the same direction as the hybrid orbital from which it hops to the NNN $p_z$ orbital. To illustrate this process, we consider the hopping from $A$ to $b_1$ via $V_9$
\begin{enumerate}
\item $(H^{z,\sigma}_{L})^\dagger$ contributes a factor proportional to $\sigma_y$, and leads to the hopping to state $A_2$
\item $H_{\sigma,\rm{NN}}^{-1}$ contributes a factor $1/V_2$, and leads to the hopping to state $a^1_2$
\item $H_{\sigma,\rm{onsite}}$ contributes a factor $V_9$, and leads to the hopping to state $a^1_1$
\item $H_{\sigma,\rm{NN}}^{-1}$ contributes a factor $1/V_3$, and leads to the hopping to state $b^1_1$
\item $(H^{z,\sigma}_{L})^\dagger$ contributes a factor proportional to $\sigma_y$, and leads to the hopping to state $b_1$.
\end{enumerate}
Since the initial and final hoppings are both proportional to $\sigma_y$, we find that the combination is proportional to $\sigma_y\cdot\sigma_y=\mathbb{I}$; hence, it does not lead to a spin-dependent hopping.
\\

Next, we would like to give some insight in the effective SOC Hamiltonians used in the six-site model. First of all, it should be noted that $\lambda_{I,4}$ does not appear in the expression for the effective parameters. This is due to the fact that we have dropped the term $H_{1,SOC}$. However, one would still expect that $\lambda_{I,4}$ should be present in the term $H_{3,SOC}$, since $H_{\beta,4}$ corresponds to the hopping from a vertex to an edge. However, the structure of $H_{3,SOC}$ immediately makes clear why this term is missing. The combination $H_{lh,SOC}\tilde{H}_{hh}^{-1}\tilde{H}_{lh}^\dagger$ corresponds from right to left to,{\it (i)} the hopping from a vertex to a NN edge,{\it (ii)} subsequently to a NNN edge, {\it (iii)} and finally back to the inital state via $H_{\beta,4}$. Furthermore, we see that both $\lambda_{ext,I}$ and $\lambda_{int, R}$ have picked up a minus sign, as follows from Eqs.~(\ref{eq2}-\ref{eq3}). Finally, the expressions can easily be related to the corresponding terms in the full $18$ site model. First of all, $\lambda_{int,I}$ is proportional to $\lambda_{I,3}$, therefore this results from the NNN hopping between orbitals at the edges, whereas $\lambda_{ext,I}$ is proportional to $\lambda_{I,2}$, which corresponds to the NNN hopping between vertices and edges. For the external Rashba SOC, we see that it originates from the Rashba SOC between the vertices. Since between two vertices there are two vertex-edge hoppings and one  edge-edge hopping, there is a term proportional to $2\lambda_{R,2}$ and a term proportional to $\lambda_{R,3}$.
\section{Calculation of Chern numbers}
The Chern numbers $C_n$ that have been calculated are defined as
\begin{align}\label{chernint}
C_n:=\frac{1}{2\pi i}\int_{BZ}d^2k\,F_{12}(k),
\end{align}
where BZ stands for Brillouin zone and $F_{12}(k)$ is the Berry curvature
\begin{align}
F_{12}(k)&:=\partial_1 A_2(k)-\partial_2 A_1(k),\nonumber\\
A_\mu(k)&:=\langle n(k)|\partial_\mu|n(k)\rangle.
\end{align}
Here, $|n(k)\rangle$ is a normalized Bloch state belonging to the nth band and $A_\mu(k)$ is the Berry connection. Note that we are implicitly assuming that the nth band is gapped, otherwise this is not a sound definition. The most elementary way to compute this number is by replacing the integral by a summation and the  derivative by some discrete differences. This procedure could be cumbersome, hence we us an alternative way of computing the Chern numbers based on lattice gauge theory [S2]. First of all, we assume that the BZ is discretized by a set of lattice points $k_l=(k_{l,x},k_{l,y})$ that are equally spaced, with spacing $a$. On this discrete set of points, we define a $U(1)$ link variable
\begin{align}
U_{\mu}(k_l)&:=\frac{\langle n(k_l)|n(k_l+\hat{\mu})\rangle}{|\langle n(k_l)|n(k_l+\hat{\mu})\rangle|}
\end{align}
where $\hat{\mu}$ is defined as the vector pointing in the direction $\mu$ with magnitude $a$. Next, we define the lattice field strength $\tilde{F}_{12}(k_l)$ as
\begin{align}
\tilde{F}_{12}(k_l):=\log{[U_1(k_l)U_2(k_l+\hat{1})U_1^{-1}(k_l+\hat{2})U^{-1}_2(k_l)]}.
\end{align}
Finally, we define the lattice Chern number as
\begin{align}\label{chern}
\tilde{C}_n:=\frac{1}{2\pi i}\sum_{k_l\in BZ}\tilde{F}_{12}(k_l).
\end{align}
It still remains to be shown that upon taking the limit $a\rightarrow 0$  we find $\tilde{C}_n=C_n$. One can easily prove that the lattice field strength is a gauge invariant quantity. As a result, we may simply assume that the normalized state $|n(k)\rangle$ varies smoothly through the BZ. Suppose now that the lattice spacing $a\ll 1$. Then, we find
\begin{align}
\tilde{F}_{12}(k)&=\log{[U_1(k)U_2(k+\hat{1})U^{-1}_1(k+\hat{2})U^{-1}_2(k)]}\nonumber\\
&\approx\log{\{U_1(k)\left[U_2(k)+a\partial_1 U_2(k)\right]U^{-1}_1(k)\left[1-a U^{-1}_1(k)\partial_2 U_1(k)\right]U^{-1}_2(k)\}}\nonumber\\
&\approx\log{[(1+a U^{-1}_2(k)\partial_1 U_2(k)-a U^{-1}_1(k)\partial_2 U_1(k))+O(a^2)]}\nonumber\\
&\approx a U^{-1}_2(k)\partial_1 U_2(k)-a U^{-1}_1(k)\partial_2 U_1(k)\nonumber\\
&\approx U_2^{-1}(k) [U_2(k+\hat{1})-U_2(k)]-1\leftrightarrow 2\nonumber\\
&\approx \left[\frac{\langle n(k+\hat{1})|n(k+\hat{1}+\hat{2})\rangle}{1+O(a)}-\frac{\langle n(k)|n(k+\hat{2})\rangle}{1+O(a)}\right](1+O(a))-1\leftrightarrow 2
\end{align}
In the fifth line, we used $|\langle n(k)|n(k+\hat{\mu})\rangle|=1+O(a)$,  as follows from the normalization. Moreover, we used $U_2^{-1}(k)\approx 1+O(a)$,  this follows from the fact that for $a=0$ we obtain $U_{\mu}(k)=1$. On the other hand, we find
 \begin{align}
 F_{12}(k)&\approx \frac{A_2(k+\hat{1})-A_2(k)}{a}-1\leftrightarrow 2\nonumber\\
 &\approx \frac{\langle n(k+\hat{1})|n(k+\hat{1}+\hat{2})\rangle-\langle n(k+\hat{1})|n(k+\hat{1})\rangle}{a^2}-\frac{\langle n(k)|n(k+\hat{2})\rangle-\langle n(k)|n(k)\rangle}{a^2}-1\leftrightarrow 2\nonumber\\
 &\approx \frac{\langle n(k+\hat{1})|n(k+\hat{1}+\hat{2})\rangle-\langle n(k)|n(k+\hat{2})\rangle}{a^2}-1\leftrightarrow 2\nonumber
 \end{align}
 Hence, $\tilde{F}_{12}(k)\approx F_{12}(k)a^2$. Plugging this into Eq.~(\ref{chern}), we find
\begin{align}
\tilde{C}_n&=\sum_{k_l\in BZ}\tilde{F}_{12}(k_l)\nonumber\\
&=\sum_{k_l\in BZ}F_{12}(k_l)a^2(1+O(a))
\end{align}
The integral in Eq.~(\ref{chernint}) may be written as a Riemann sum, $C_n=\lim_{a\rightarrow 0}\sum_{k_l\in BZ}a^2 F_{12}(k_l)/(2\pi i)$. We then find  $C_n=\tilde{C}_n$ upon taking the limit $a\rightarrow 0$. By using this method, we have obtained  the Chern numbers, as shown in the main text (see also Fig.\ 3 therein).
\section{Method for deriving effective Hamiltonians}\label{efff}
Here, we outline the method used throughout the paper to derive effective Hamiltonians. Suppose we are given a system such that we can split the spinor $\Psi$ into a high-energy component $\Psi_h$ and a low-energy component $\Psi_l$. Then, we may write the Hamiltonian matrix in a corresponding block form
\begin{align}\label{0}
H(k)&=\begin{pmatrix}
H_{ll}(k)&H_{lh}(k)\\
H_{lh}^\dagger(k)&H_{hh}(k)
\end{pmatrix}.
\end{align}
Using this decomposition, the Schr\"odinger equation reads
\begin{align}
E\Psi_l(k)&=H_{ll}(k)\Psi_l(k)+H_{lh}(k)\Psi_h(k)\label{1}\\
E\Psi_h(k)&=H_{lh}^\dagger(k) \Psi_l(k)+H_{hh}(k)\Psi_h(k).\label{2}
\end{align}
We can then use Eq. (\ref{2}) to eliminate $\Psi_h$ in Eq. (\ref{1}). Since $H_{lh}^\dagger\Psi_l=(-H_{hh}+E)\Psi_h$, up to first order in $E$ we obtain $\Psi_h=-H_{hh}^{-1}(1+EH_{hh}^{-1})H_{lh}^\dagger\Psi_l$. Therefore, Eq. (\ref{1}) reduces to
\begin{eqnarray}\label{4}
(H_{ll}-H_{lh}H_{hh}^{-1}H_{lh}^\dagger)\Psi_l&=E(\mathbb{I}+H_{lh}H_{hh}^{-2}H_{lh}^\dagger)\Psi_l.
\end{eqnarray}
If we now introduce $S=\mathbb{I}+H_{lh}H_{hh}^{-2}H_{lh}^\dagger$, and define $\phi=S^{1/2}\Psi_l$, we find the eigenvalue equation
\begin{align}\label{3}
(H_{ll}-H_{lh}H_{hh}^{-1}H_{lh}^\dagger)S^{-1/2}\phi&=E S^{1/2} \phi.
\end{align}
By multiplying Eq.~(\ref{3}) on both sides with $S^{-1/2}$ we find
\begin{align}
H_{\rm{eff}}\phi&=E\phi,
\end{align}
with $H_{\rm{eff}}$ given by
\begin{align}
H_{\rm{eff}}&=S^{-1/2}(H_{ll}-H_{lh}H_{hh}^{-1}H_{lh}^\dagger)S^{-1/2}.
\end{align}

\underline{\bf Supporting References:}\\

{[S1]} \textrm{S. Konschuh, M. Gmitra, and J. Fabian},  { Phys. Rev. B} {\bf 82}, 245412 (2010).\\

{[S2]} \textrm{T. Fukui, Y. Hatsugai, and H. Suzuki}, { J. Phys. Soc. Jpn.} {\bf 74}, 1674 (2005).

\end{widetext}

\end{document}